\documentclass[prb,aps,twocolumn,floatfix,a4paper]{revtex4}

\usepackage{amsmath,amssymb,amsfonts}
\usepackage{bm}\let\vec\bm
\usepackage{graphicx}
\usepackage{mathrsfs}
\renewcommand{\Im}{\mathrm{Im}}
\renewcommand{\Re}{\mathrm{Re}}
\newcommand{\figwidth}{.8\columnwidth}
\newcommand{\fref}[1]{Fig.~\ref{#1}}

\begin{document}

\title{Critical properties and Bose Einstein Condensation in dimer spin systems}

\author{E. Orignac}
\affiliation{LPENSL CNRS UMR 5672  \\ 46 All\'ee d'Italie,  69364 Lyon
  Cedex 07, France }

\author{R. Citro}
\affiliation{ Dipartimento di Fisica ``E. R. Caianiello'' and
Unit{\`a} C.N.I.S.M. di Salerno\\ Universit{\`a} degli Studi di
Salerno, Via S. Allende, I-84081 Baronissi (Sa), Italy}

\author{T. Giamarchi}
\affiliation{DPMC-MaNEP, University of Geneva, 24 quai Ernest
Ansermet, 1211 Geneva 4, Switzerland}

\begin{abstract}
We analyze the spin relaxation time $1/T_1$ for a system made of weakly coupled one dimensional ladders.
This system allows to probe the dimensional crossover between a Luttinger liquid and a Bose-Einstein condensate
of magnons. We obtain the temperature dependence of $1/T_1$ in the various dimensional regimes, and discuss the experimental
consequences.
\end{abstract}
\date{\today}
\maketitle

Bose-Einstein condensation (BEC) is a particularly spectacular
manifestation of quantum physics and the indiscernibility of the
particles. Traditional systems where to observe such a phenomenon
are bosonic systems such as \cite{tilley_book_superfluidity} Helium 4
or more recently cold atomic
gases \cite{pitaevskii_becbook}. However BEC is also present in
many other condensed matter systems. Superconductivity can be viewed
as a BEC of Cooper pairs and
BEC has also long been sought in excitonic systems \cite{kasprzak_BEC_polaritons}.
Because spin-1/2 operators can be
faithfully represented by hard core bosons \cite{auerbach_book_spins}, spin systems are also very natural
systems where to look for such a phenomenon \cite{batyev_spins_field,affleck_spin1_field,giamarchi_coupled_ladders}
and various applications have thus been investigated.
However, in practice, both the experimentally large exchange
constants and, on a more theoretical level the hard core
constraint, which corresponds to very strong repulsion of the
bosons, restrict considerably the usefulness of these concepts for
simple regular spin systems. In the extreme case of one dimension
it is in fact much more fruitful to think of spins-1/2 in terms of
spinless fermions \cite{giamarchi_book_1d} than in terms of bosons.

Nevertheless, it has been realized \cite{giamarchi_coupled_ladders} that the
BEC physics is extremely useful to describe the
spin systems made of coupled spin-1/2 dimers.
In such systems the
application of a magnetic field induces a zero temperature
quantum phase transition
(QPT) between a spin-gap phase with zero magnetization
and a phase where some of the dimers are polarized in
a triplet state giving a nonzero magnetization.
This transition has been shown
\cite{giamarchi_coupled_ladders} to be in the universality class
of BEC, where the bosons represent the dimers in the  triplet
state. The BEC order parameter is then the staggered magnetization
in the plane orthogonal to the applied field.
These predictions have been spectacularly
confirmed by magnetization \cite{nikuni_bec_tlcucl} and neutron
\cite{ruegg_bec_tlcucl} measurements on TlCuCl$_3$ and related dimer
compounds. The active discovery and analysis of new compounds in
which this phenomenon could be observed thus triggered further
theoretical
\cite{wessel00_bec_magnons,wessel01_spinliquid_bec,nohadani04_scaling,nohadini05_qpt,matsumoto02_tlcucl3,%
misguich04_tlcucl3,matsumoto04_tlcucl3,sirker04_bec_dm,ng06_magnon_bec} and experimental work \cite{oosawa02_kcucl3,sherman03_tlcucl3,grenier04_cr3cr2br9,radu05_cs2cucl4,jaime04_bacu2si2o6,%
sebastian05_han_pigment,paduan-filho04_nicl2,stone05_organic_qcp,fujii05_bipnnbno,toda05_csfecl3,tsuji05_pbni2v2o8}
on this magnon BEC.
\begin{figure}
\includegraphics[width=\figwidth]{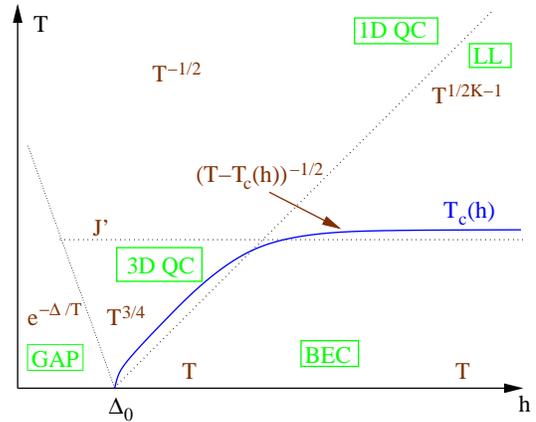}
\caption{\label{fig:nmrt1} Summary of the temperature dependence of 1/T1.
$T_c(h)$ represents the BEC condensation temperature. $\Delta_0$ is the gap in the absence of magnetic field.
The temperature dependence is indicated for the various regimes (see text).}
\end{figure}

In connection with the above mentioned magnon BEC, an important
question is to understand how the dimensional crossover, between the
one dimensional and the three dimensional behavior can occur in
systems made of weakly coupled ladders. From a theoretical point of
view, in such systems, at  temperatures  much larger than the
inter-ladder coupling, the system can be viewed as a collection of one
dimensional ladders, with the very peculiar associated one dimensional
behavior
\cite{sachdev_qaf_magfield,chitra_spinchains_field,giamarchi_coupled_ladders,furusaki_dynamical_ladder},
more fermionic in nature, and in the universality class of Luttinger
liquids at high field\cite{giamarchi_book_1d}. At lower temperature,
interladder coupling cannot be ignored anymore, and the system falls
into  the universality class of magnon BEC
condensates. From the experimental
point of view,  such a change of regime between the one
dimensional limit and the three dimensional one is relevant for
compounds such as BPCP \cite{watson_bpcb} that are a very good
realization of weakly coupled two legs ladders, with strong dimers
along the ladder rungs. To probe these different regimes and the
dimensional crossover, it is crucial to understand the spin
dynamics as a function of temperature and magnetic field. An important
probe in that respect is provided by Nuclear Magnetic Resonance (NMR)
measurements.  Quantities such as the relaxation rate $1/T_1$ have been
computed in the condensed phase and in the one dimensional Luttinger
liquid regime
\cite{chitra_spinchains_field,giamarchi_coupled_ladders}. However,
predictions for their behavior close to the BEC quantum critical point or
in the other one-dimensional regimes
are still lacking, although some results for two
dimensional systems have recently appeared
\cite{sachdev_2d_qcp_spins}.

In this paper we thus calculate the NMR relaxation rate $1/T_1$ as a
function of temperature and magnetic field
in a system made of weakly three-dimensionally coupled ladders, in the
one-dimensional and in the three-dimensional regime.
Our findings are summarized in Fig.~\ref{fig:nmrt1}

We consider a system of coupled one dimensional two leg ladders described by the Hamiltonian
\begin{multline} \label{eq:ladderham}
 H = J \sum_{\alpha,i,l=1,2} \vec{S}_{\alpha,i,l} \cdot \vec{S}_{\alpha,i+e_x,l}
 + J_\perp \sum_{\alpha,i} \vec{S}_{\alpha,i,1} \cdot \vec{S}_{\alpha,i,2} \\
 - h \sum_{\alpha,i,l=1,2}
 S^z_{\alpha,i,l} + H'
\end{multline}
where $\alpha$ is the ladder index, $l=1,2$ denote the two legs of the ladder, $e_x$ is the direction parallel to the ladder leg and $h$ the applied
magnetic field. $H'$ is the interladder exchange term.
We take here the simple form for $H'$ :
\begin{equation}
 H' = J' \sum_{\alpha,i} \vec{S}_{\alpha,i,2} \cdot \vec{S}_{\alpha,i+e_y,1} + J'' \sum_{\alpha,i,l} \vec{S}_{\alpha,i,l} \cdot \vec{S}_{\alpha,i+e_z,l}
\end{equation}
where $e_y$ is the direction of the rungs, and $e_z$ is the direction orthogonal
to the ladder planes. We take for simplicity $J'=J''$, generalization is immediate.
The NMR relaxation rate $T_1$ can be expressed \cite{moriya_nmr} as a
function of the retarded spin-spin correlation function. We
concentrate in this paper on the contribution of the transverse local
spin-spin correlation,
\begin{equation}\label{lin-rep}
 \frac1{T_1} = T \lim_{\omega\to 0} \sum_{q} \frac{\Im \langle
   S^+(q,\omega) S^-(q,\omega)\rangle}\omega,
\end{equation}
as one can show that the longitudinal contribution
$\langle S^z S^z \rangle$ gives
only subdominant corrections as far as the
temperature dependence is concerned \cite{tobepublished}.

When the temperature is larger than the interladder coupling $J'$ the
system can be essentially described by a set of independent ladders.
This one-dimensional regime is well understood
\cite{sachdev_qaf_magfield,chitra_spinchains_field,giamarchi_coupled_ladders}.
For fields $h$ lower than the spin gap of the ladder $\Delta_0$, and
at sufficiently low temperature, the system remains in the spin gap
state, and $1/T_1$ has an activated form \cite{kishine_nmr_ladder}.
For fields $h>\Delta_0$ , and such that $h-\Delta_0 \gg T$,
the system is in a Luttinger liquid regime
\cite{chitra_spinchains_field,giamarchi_coupled_ladders,furusaki_dynamical_ladder}
and one finds $1/T_1 \sim T^{1/(2K)-1}$ for $h>\Delta_0$, with $K$ the
Luttinger exponent. Close to the critical field the ladder system can
be mapped to a two state system for which only the singlet and lowest
triplet are important
\cite{tachiki_copper_nitrate,totsuka_ladder_strongcoupling,chaboussant_ladder_strongcoupling,mila_field,giamarchi_coupled_ladders}.
This allows to use a pseudo spin 1/2 representation. For high temperatures, $T\gg
|h-\Delta_0|$, the system is in
a universal quantum critical regime \cite{sachdev_qaf_magfield} with a dynamical exponent $z=2$.
There, $1/T_1$ varies as a power law with
a universal exponent which can be determined by a
scaling argument. Indeed, scaling requires a
local spin correlation function  of
the form $T^{-1/2} f(\omega/T)$ in one dimension. The
requirement that $f$ be analytical in $\omega$ for $T>0$ with a vanishing
imaginary part for $\omega=0$ implies $T/\omega\times
\mathrm{Im}f(\omega/T) \to \mathrm{Ct.}$,  leading to $1/T_1 \sim
T^{-1/2}$. We note that since at the commensurate-incommensurate
transition induced by the magnetic field, one predicts
\cite{chitra_spinchains_field,giamarchi_coupled_ladders} that $K=1$,
the behavior of $1/T_1$ is continuous at the crossover from the
Luttinger liquid to the $z=2$ quantum critical regime. The exponent simply
becomes constant once the $z=2$ quantum critical regime has been reached.
Such a behavior is shown in \fref{fig:nmrt1}.

\begin{figure}[htbp]
  \centering
  \includegraphics[width=5cm]{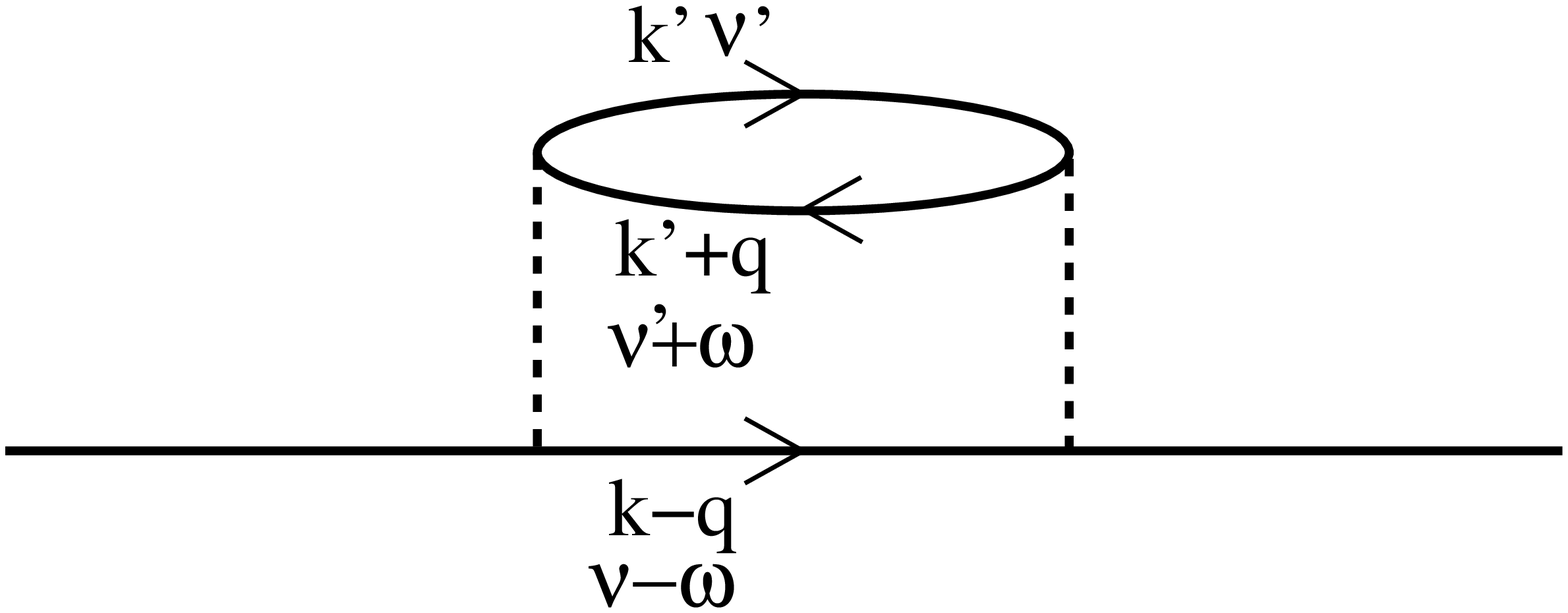}
  \includegraphics[width=5cm]{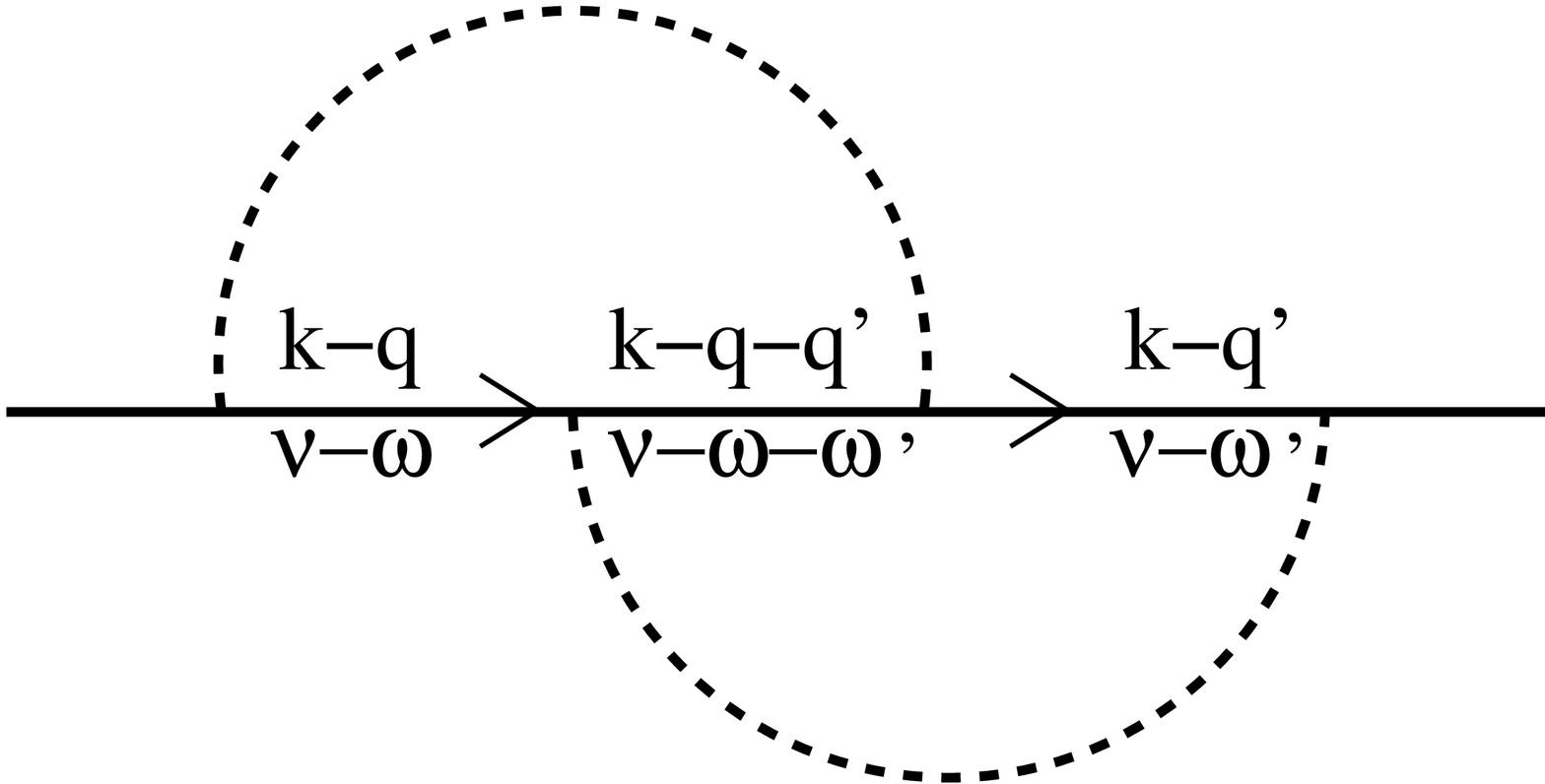}
  \caption{Skeleton self-energy graphs at second order.
  The dotted line represents the interaction and the straight line
  the magnon propagator. The magnon propagator is computed with the first order self energy.
  Because of the low density limit close to the quantum critical point this expansion is well behaved.}
  \label{fig:graphs}
\end{figure}

At low temperature the system enters a three dimensional regime, where
the interladder coupling must be considered from the start.  Two
distinct regimes occur depending on the magnetic field. If $h-\Delta_0
\gg J'$, then the system can be viewed as a collection of one
dimensional ladders, with a fully developed Luttinger regime for each
ladder. The coupling between the ladders then induces a three
dimensional ordering for the magnetization in the XY plane, in the
same universality class than a Bose-Einstein condensation
\cite{giamarchi_coupled_ladders}. The one dimensional fluctuations
strongly affect the three dimensional ordering temperature to lead to
$T_{BEC}=J(J'/J)^{1/(2-2K)}$.  This regime can essentially be
described by an RPA treatment of the one dimensional ladders.  
If $h-\Delta_0 \ll J'$ then the coupling between the ladder is too strong for the ladders to have a
one dimensional regime. One must then view the system as an
anisotropic, but three dimensional, spin system. Note that this regime is universal and depends only
on the presence of dimers in a three dimensional environment and not on whether a ladder structure exists.
The results we obtain in this paper for the 3D quantum critical regime thus also apply to compounds for which
$J_\perp \gg (J,J')$ without any specific relation between $J$ and $J'$.
A useful representation
\cite{giamarchi_coupled_ladders,nikuni_bec_tlcucl} is to introduce a
boson excitation for each rung in the ladder.  The presence (resp.
absence) of the boson indicates that the rung is the triplet (resp.
singlet) state.  In the regime with three-dimensional coupling, and
above the BEC condensation temperature, we can describe the coupled
dimer system (\ref{eq:ladderham}) using the following effective
Hamiltonian
\cite{giamarchi_coupled_ladders,nikuni_bec_tlcucl,misguich04_tlcucl3}:
\begin{eqnarray}
  H=\sum_k (\epsilon(k)-h) b^\dagger_k b_k + \frac{g}{L^3}
  \sum_{k_1,k_2,q} b^\dagger_{k_1+q} b^\dagger_{k_2-q} b_{k_2} b_{k_1},
\end{eqnarray}
where $\epsilon(k)=\Delta_0+\frac{k^2}{2m}$ is the dispersion of
the $S^z=1$ gapped magnons, $b^\dagger_k$ creates a magnon
excitation with momentum $k$ and $g$ is an effective repulsion
between the magnons. Since the spin operator
$S^+$  is proportional to the magnon creation operator
$b^\dagger$, the  correlation function in~(\ref{lin-rep}) can be
obtained from the Matsubara Green's function of the magnon
operator. Above the BEC temperature, anomalous propagators are vanishing,
and this Matsubara Green's function is obtained as a function of
the magnon self-energy $\Sigma(k,\omega+i0)$ from the usual Dyson
equation. We therefore have:
\begin{widetext}
  \begin{eqnarray} \label{eq:nmr-rate}
  \frac 1 {T_{1}} \sim \lim_{\omega \to 0} \frac T \omega \int \frac{d^3 k}{(2\pi)^3}
  \frac{\Im \Sigma(k,\omega)} {(\omega +h -\epsilon(k) -\Re
  \Sigma(k,\omega))^2+\Im^2 \Sigma(k,\omega)}.
\end{eqnarray}
\end{widetext}
Just above the BEC temperature, the magnon density is small, making a low
density expansion applicable. At first order
\cite{nikuni_bec_tlcucl} only Hartree-Fock diagrams are present,
however they
 do not contribute to the imaginary part of the self-energy. The
first non-trivial contribution to the imaginary part of the magnon
self-energy comes from the second order graphs shown on
Fig.~\ref{fig:graphs}.
In order to evaluate the expression (\ref{eq:nmr-rate}), we use
the fact that when $\omega \to 0$, $T \Im \Sigma(k,\omega)/\omega$ has
a finite limit. This implies that the denominator in Eq.~
(\ref{eq:nmr-rate})  is dominated by the
contribution coming from the real part of the Green's function.
Since the contribution coming from the denominator  is strongly peaked
near $k=0$ when $h$ is near $\Delta_0$, the
dominant contribution in Eq.~(\ref{eq:nmr-rate})
comes from the small momenta. The
width in momentum of the expression in the numerator
\cite{tobepublished} being of order  $T^{2/3}$, it is less strongly
peaked near $k=0$  than the contribution from  the denominator for
$T\to 0$.  This justifies the replacement  in the
numerator, of $T \Im \Sigma(k,\omega)/\omega$ by its value for
$k=0,\omega=0$, yielding $1/T_1 \sim T [m^3/2(\Delta-h)]^{1/2} \lim_{\omega \to 0}
\Im \Sigma(k=0,\omega)/2\pi \omega$, with $\Delta$  the
effective gap containing the Hartree-Fock renormalization of $\Delta_0$.

Following the same steps as in Ref.~\onlinecite{sachdev06_bose2d},  the
graphs of Fig.~\ref{fig:graphs} are evaluated in the limit $k\to 0$ as:
\begin{widetext}
\begin{equation}\label{master}
   \Im \Sigma(k=0,\omega)=\frac{m^3 g^2 T}{4\pi^3} (1-e^{- \omega/T})\int_0^\infty
   d\epsilon_1 \frac{e^{-(\Delta
   +\epsilon_1-\omega)/T}}{(1-e^{-(\Delta +\epsilon_1-\omega)/T})
   (1-e^{-(\Delta +\epsilon_1)/T})}
   \ln\left(1+\frac{1-e^{-(\epsilon_1+\Delta -\omega)/T}}{e^{\frac 1 T \left(\frac{(\omega-\Delta)^2}{4\epsilon_1}+\Delta\right)}-1}\right),
\end{equation}
\end{widetext}
In the vicinity of the quantum critical point, $\Delta \ll T$ and the
integral~(\ref{master}) yields:
\begin{eqnarray}\label{eq:T1-qcp}
  \frac 1 {T_1} \sim m^{9/2} {g^2}
  \frac {T^3} {(\Delta-h)^{3/2}}
\end{eqnarray}
Let us note that by a power-counting argument \cite{baym01_dilute_bec},
we expect  the real part of the $n$-th  order correction  to the
self-energy to behave as $g^n T^{1+n/2}$. This implies that at low
temperature the
Hartree-Fock contribution dominates so that  $\Delta(T)-h \sim
g T^{3/2}$ in the 3D quantum critical regime.  Therefore, in the 3D
quantum critical regime:
\begin{equation}
 1/{T_1} \sim T^{3/4} .
\end{equation}
In the BEC regime, we find instead that $\Delta(T)-h \propto
(T-T_{BEC}(h))$ and $1/T_1\propto (T-T_{BEC}(h))^{-3/2}$. This
behavior is valid when $T-T_{BEC}(h) \gg |\Delta_0-h|$ where
$\Delta_0$ is the bare gap. This result breaks down when $T$ is
sufficiently close to $T_{BEC}(h)$. Then, one has the mean-field
behavior $1/T_1\sim
(T-T_{BEC}(h))^{-1/2}$. Below the BEC condensation temperature, due to
the presence of a massless Goldstone mode associated with phase
fluctuations, one can show \cite{giamarchi_coupled_ladders} that $1/T_1\sim T$.
At the other extreme, in the gapped regime, $|\Delta-h |\gg T$ and
evaluating~(\ref{master}) yields
$1/{T_{1}} \propto T^{3/2} e^{-3 \Delta/T}$.  Our findings are
summarized on Fig.~\ref{fig:nmrt1} for all the different regimes.

The predictions of Fig.~\ref{fig:nmrt1} are directly testable on
coupled ladder systems.
For systems made of weakly coupled ladders such as BPCB
\cite{watson_bpcb}, given the
accessible magnetic fields and with the condition $J' \ll J$, the one
dimensional regime is probably the easiest to probe.
Thus the clearest test is the
$T^{-1/2}$ divergence in the quantum critical regime.
In principle, one could  also observe
the gradual change of the exponent as one moves inside the Luttinger
regime towards quantum criticality.
However, since the  the exponent varies
between $1/2$ and $1/3$, this change is rather small and this
effect is likely to be hard to observe.
Unless one is very close to the quantum critical
point the pure one dimensional divergence will merge with the
divergence as $1/(T-T_c)^{1/2}$ at the finite temperature BEC
transition due to the three dimensional coupling. For systems for
which one is able to probe both the one dimensional and the three
dimensional regimes, a clear signature of the dimensional crossover
will be the divergence of $1/T_1$ in the one dimensional regime
followed by a drop of  $1/T_1$ in the quantum critical 3D regime.
Note that going towards the BEC regime will reveal a more complex
behavior where the initial drop when entering the 3D quantum critical
regime will be followed by a divergence at $T=T_{BEC}$ and then by
another drop inside the BEC phase.
Finally more isotropic systems would show a broad
3D regime, allowing to test the quantum critical prediction $T^{3/4}$
in the quantum critical
regime or $T$ in the BEC one.
Unfortunately in TlCuCl$_3$ the 3D quantum critical regime is
affected by a simultaneous lattice distortion
\cite{vyaselev_tlcucl_nmr}
and one must find other candidates to probe for this regime.
An interesting one might be provided by hydrated copper nitrate
\cite{amaya_copper_nitrate,vantol_hydrated_nmr}.

We thank C. Berthier for many illuminating discussions.
This work was supported by the Swiss National Science Foundation through
MaNEP and Division II.
E. O. and R. C. acknowledge hospitality and support from the University of Geneva.  E. O. acknowledges hospitality and support from
the University of Salerno and CNISM during his stay at the University
of Salerno. R. C. acknowledges hospitality and support from the ENS of
Lyon.


\end{document}